\documentclass[12pt]{iopart}
\usepackage{graphicx}

\begin{document}

\title[S. Damjanovic: Thermal dileptons at SPS energies]{Thermal dileptons at SPS energies}

\author{S Damjanovic (for the NA60 Collaboration)}

\address{CERN, 1211 Geneva 23, Switzerland}
\ead{sdamjano@cern.ch}
\begin{abstract}
Clear signs of excess dileptons above the known sources were found at
the SPS since long. However, a real clarification of these
observations was only recently achieved by NA60, measuring dimuons
with unprecedented precision in 158A GeV \, In-In collisions. The
excess mass spectrum in the region $M$$<$1 GeV is consistent with a
dominant contribution from $\pi^{+}\pi^{-} \rightarrow \rho
\rightarrow \mu^{+}\mu^{-}$ annihilation. The associated $\rho$
spectral function shows a strong broadening, but essentially no shift
in mass. In the region $M$$>$1 GeV, the excess is found to be prompt,
not due to enhanced charm production. The inverse slope parameter
$T_\mathrm{eff}$ associated with the transverse momentum spectra rises
with mass up to the $\rho$, followed by a sudden decline above. While
the initial rise, coupled to a hierarchy in hadron freeze-out, points
to radial flow of a hadronic decay source, the decline above signals a
transition to a low-flow source, presumably of partonic origin. The
mass spectra show at low transverse momenta the steep rise towards low
masses characteristic for Planck-like radiation. The polarization of
the excess referred to the Collins Soper frame is found to be
isotropic. All observations are consistent with the interpretation of
the excess as thermal radiation.
\end{abstract}


Dileptons are particularly attractive to study the hot and dense QCD
matter formed in high-energy nuclear collisions. In contrast to
hadrons, they directly probe the entire space-time evolution of the
expanding system, escaping freely without final-state interactions. At
low masses $M$$<$1 GeV (LMR), thermal dilepton production is mediated
by the broad vector meson $\rho$ (770) in the hadronic phase. Due to
its strong coupling to the $\pi\pi$ channel and the short life time of
only 1.3 fm/c, ``in-medium'' modifications of its mass and width close
to the QCD phase boundary have since long been considered as the prime
signature for {\it chiral symmetry
restoration}~\cite{Pisarski:mq,Rapp:1995zy,Brown:kk}. At intermediate
masses $M$$>$1 GeV (IMR), it has been controversial up to today
whether thermal dileptons are dominantly produced in the earlier
partonic or in the hadronic phase, based here on hadronic processes other
than $\pi\pi$ annihilation. Originally, thermal emission from the
early phase was considered as a prime probe of {\it
deconfinement}~\cite{McLerran:1984ay,Kajantie:1986}.

\begin{figure*}[]
\begin{center}
\includegraphics*[width=0.41\textwidth,clip, bb =3 32 535 535]{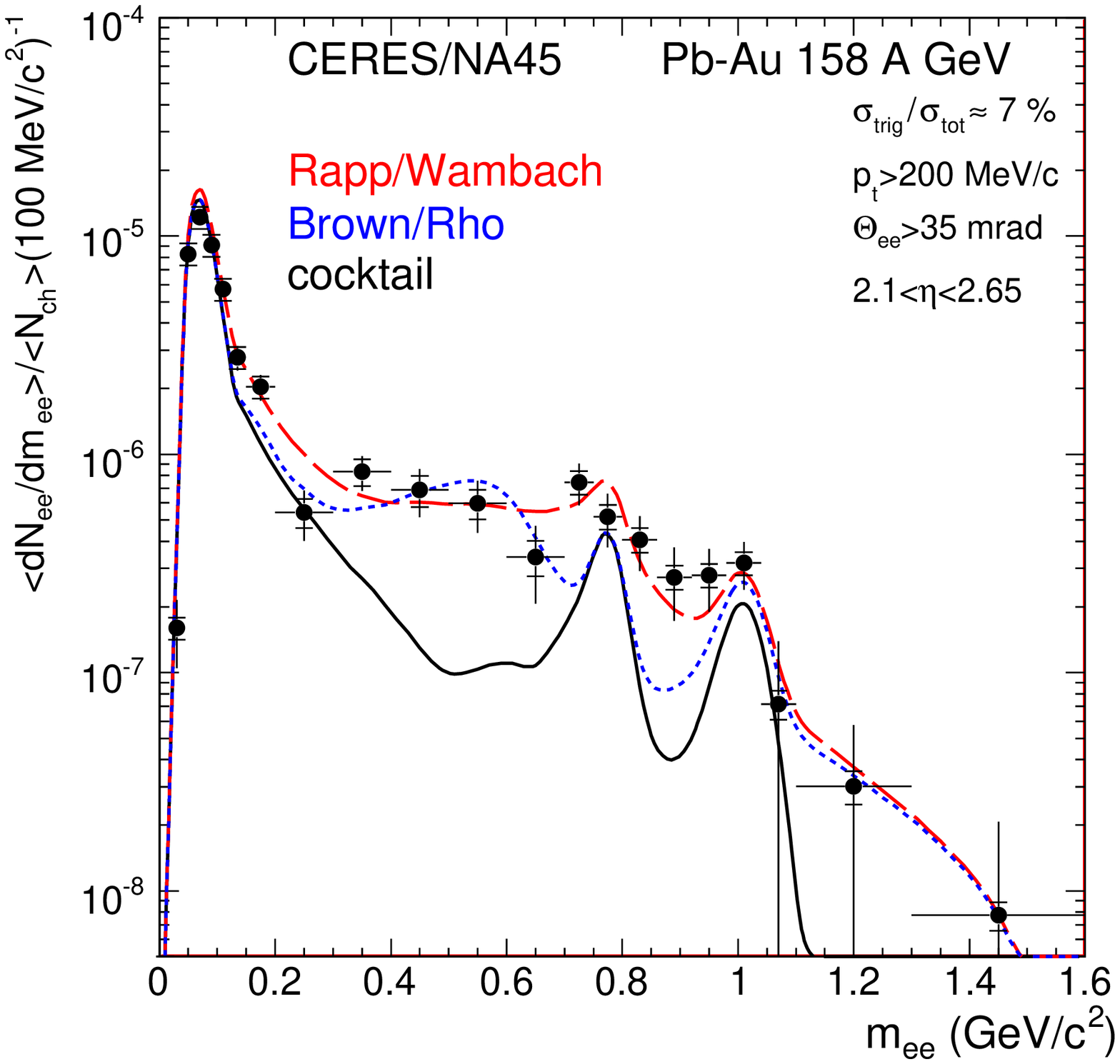}
\includegraphics*[width=0.39\textwidth]{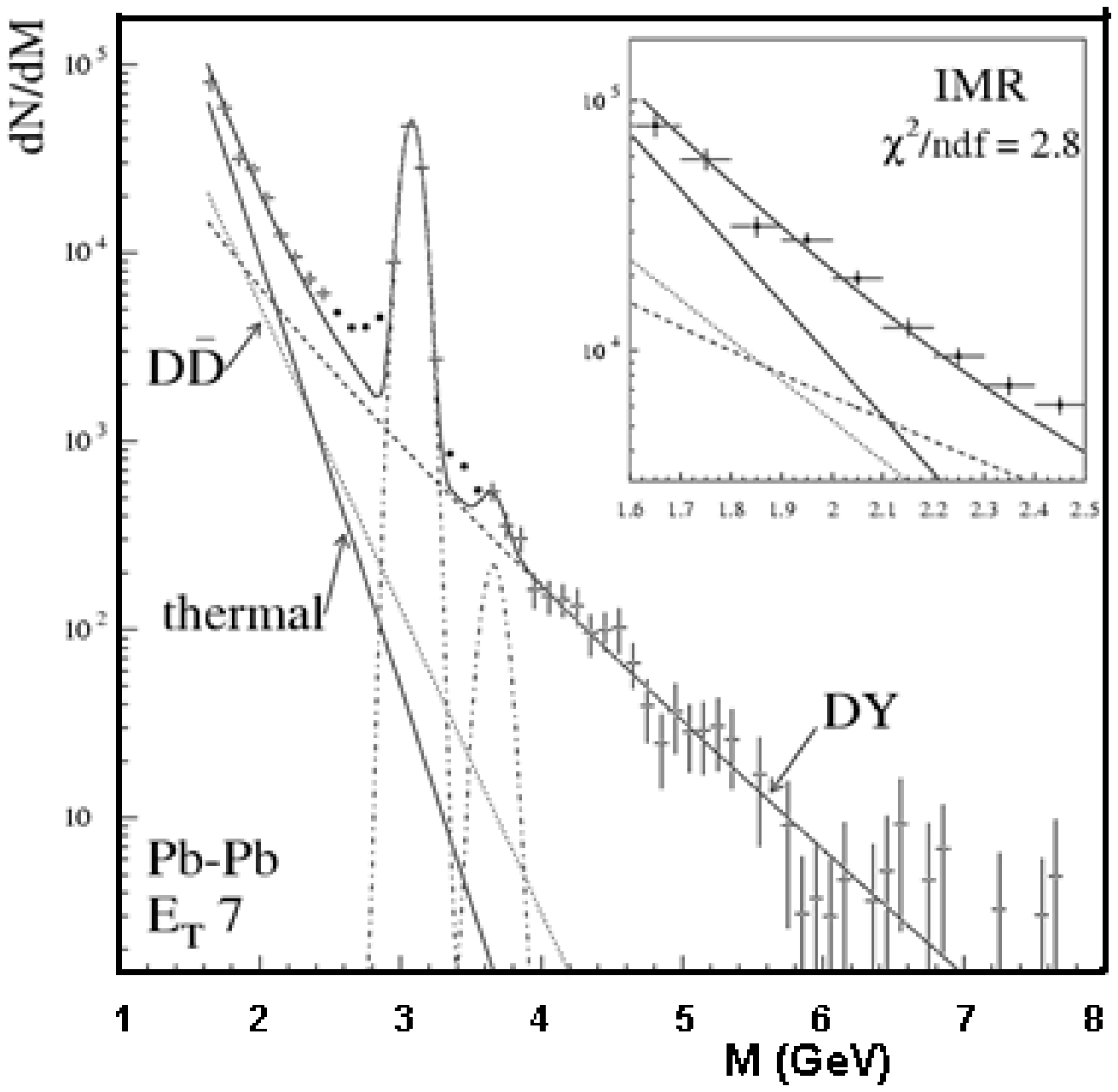}
\caption{Excess dileptons seen in previous SPS experiments by
CERES~\cite{CERES:2008} (LMR, left) and NA50~\cite{Abreu:2002rm} (IMR,
right). For the former, there is now some sensitivity (around 0.9 GeV)
to specific theoretical predictions.}
   \label{fig1}
\end{center}
\end{figure*}

Experimentally, it took more than a decade to master the challenges of
very rare signals and enormous combinatorial backgrounds. The first
clear signs of an excess of dileptons above the known decay sources at
SPS energies were obtained by CERES~\cite{Agakichiev:1995xb} for
$M$$<$1 GeV, NA38/NA50~\cite{Abreu:2002rm} for $M$$>$1 GeV and by
HELIOS-3~\cite{Masera:1995ck} for both mass regions
(see~\cite{Specht:2007ez} for a short recent review including the
preceding $pp$ era and the theoretical milestones). The final status
reached by CERES~\cite{CERES:2008} and NA50~\cite{Abreu:2002rm} is
illustrated in Fig.~\ref{fig1}.
 The sole existence of an excess gave a
strong boost to theory, with hundreds of publications. In the LMR
region, $\pi\pi$ annihilation with regeneration and strong in-medium
modifications of the intermediate $\rho$ during the fireball expansion
emerged as the dominant source. However, the data quality in terms of
statistics and mass resolution remained largely insufficient for a
precise assessment for the in-medium spectral properties of the
$\rho$. In the IMR region, thermal sources or enhanced charm production
could account for the excess equally well, but that ambiguity could
not be resolved, nor could the nature of the thermal sources be
clarified.

A big step forward in technology, leading to completely new standards
of the data quality in this field, has recently been achieved by NA60,
a third-generation experiment built specifically to follow up the open
issues addressed above~\cite{Gluca:2005}. Initial results on mass and
transverse momentum spectra of the excess dimuons have already been
published~\cite{Arnaldi:2007ru,Arnaldi:2006jq}. This paper shortly
reviews these results, but also reports on further aspects associated
with the centrality dependence, polarization, acceptance-corrected
mass spectra and absolute normalization.

\begin{figure*}[]
\begin{center}
\includegraphics*[width=0.43\textwidth]{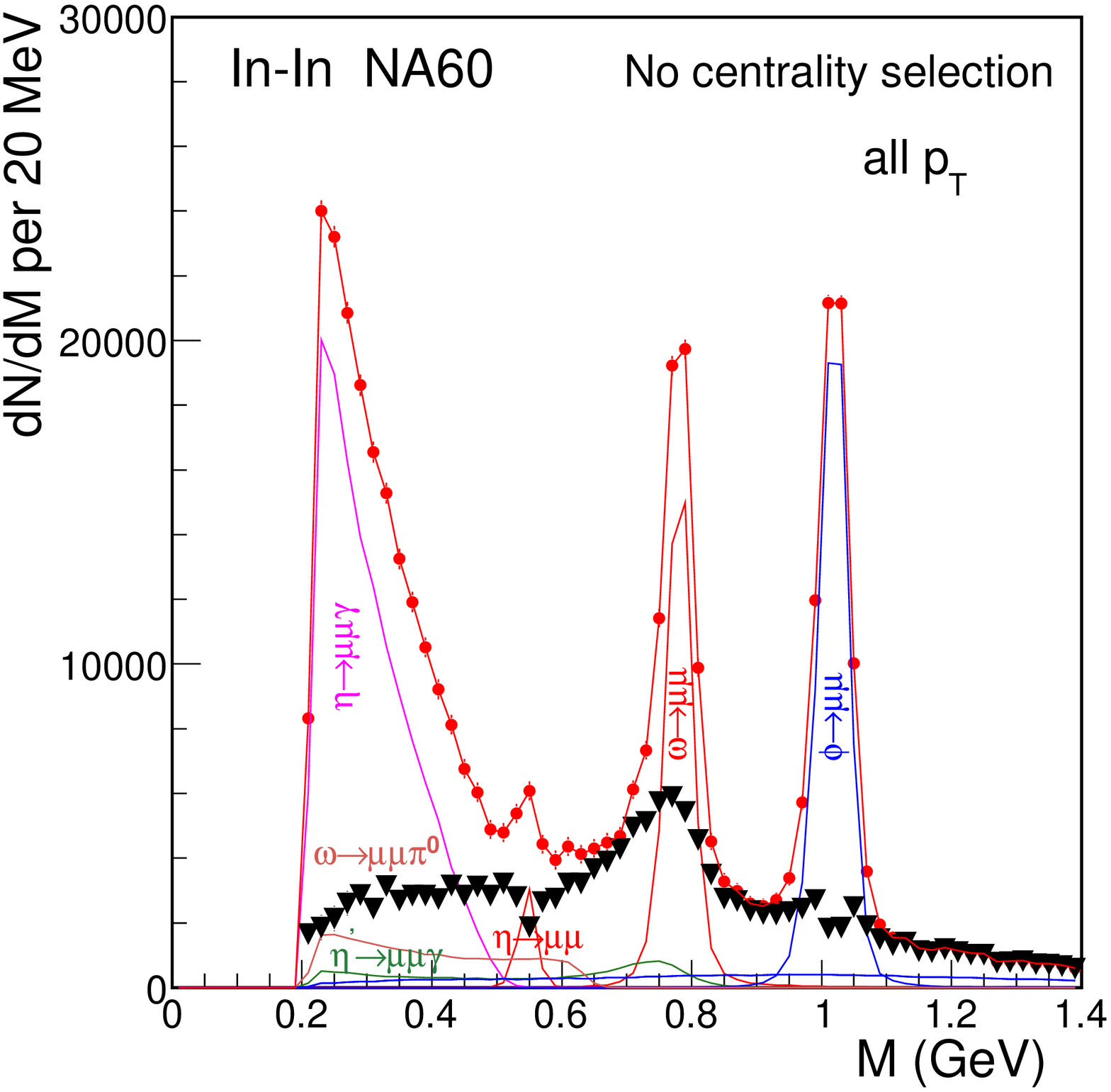}
\includegraphics*[width=0.43\textwidth]{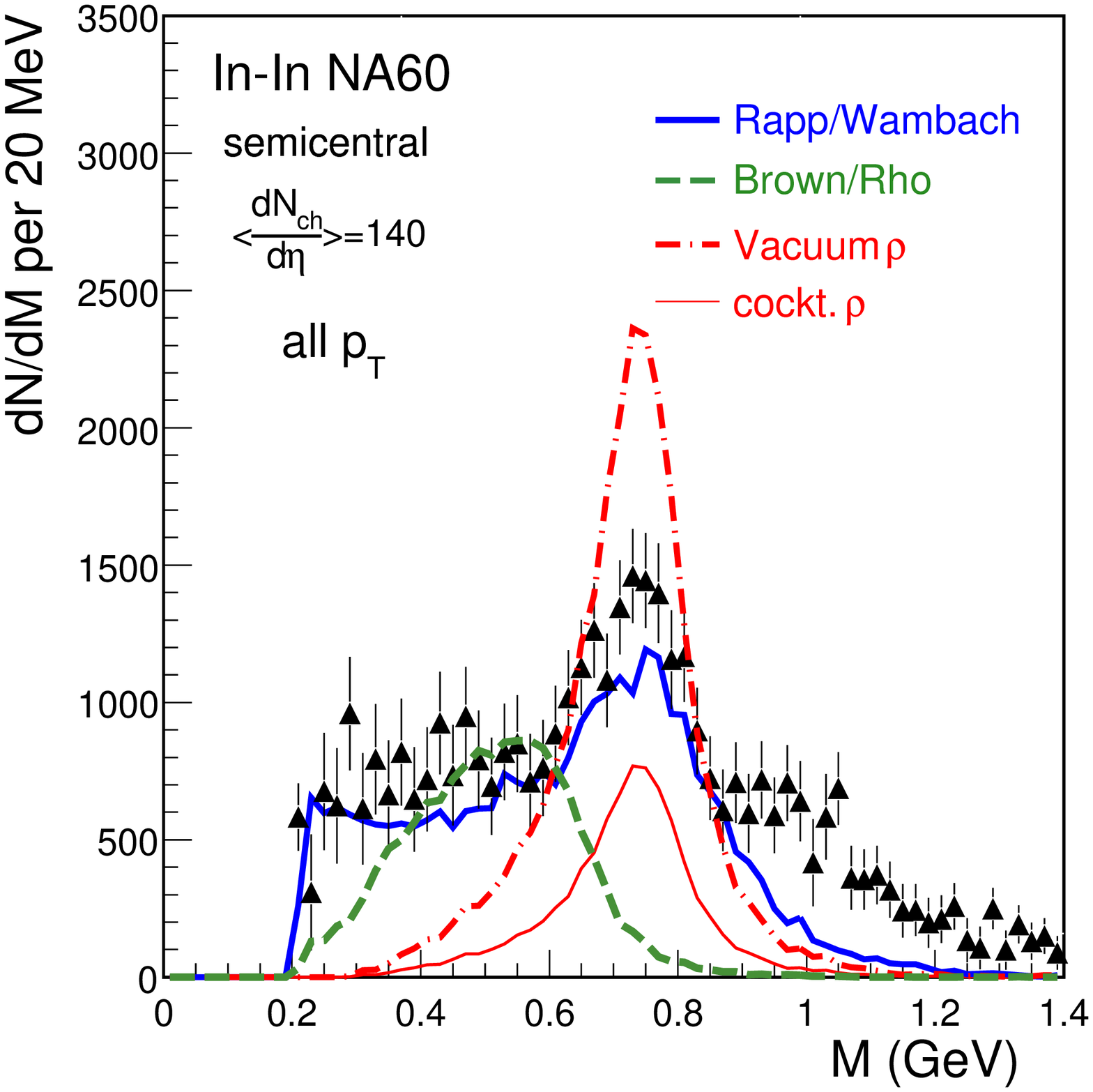}
\caption{Background-subtracted mass spectrum before (dots) and after
subtraction of the known decay sources (triangles). Right: Excess dimuons
compared to theoretical predictions, keeping the original
normalization in absolute terms~\cite{Rapp:2005pr}.}
   \label{fig2}
\end{center}
\end{figure*}

Fig.~\ref{fig2} (left) shows the centrality-integrated net dimuon mass
spectrum for 158A GeV In-In collisions in the LMR region. The narrow
vector mesons $\omega$ and $\phi$ are completely resolved; the mass
resolution at the $\omega$ is 20 MeV. The peripheral data can be
completely described by the electromagnetic decays of neutral
mesons~\cite{Arnaldi:2007ru,Damjanovic:2006bd}. This is not true for
the more central data as plotted in Fig.~\ref{fig2}, due to the
existence of a strong excess of pairs. The high data quality of NA60
allows to isolate this excess with {\it a priori unknown
characteristics} without any fits: the cocktail of the decay sources
is subtracted from the total data using {\it local} criteria, which
are solely based on the mass distribution itself. The $\rho$ is not
subtracted. The excess resulting from this difference formation is
illustrated in the same figure
(see~\cite{Arnaldi:2007ru,Arnaldi:2006jq,Damjanovic:2006bd} for
details and error discussion).

The common features of the excess mass spectra can be recognized in
Fig.~\ref{fig2} (right). A peaked structure is always seen, residing
on a broad continuum with a yield strongly increasing with centrality
(see below), but remaining essentially centered around the nominal
$\rho$ pole~\cite{Damjanovic:2006bd}. Without any acceptance
correction and $p_{T}$ selection, the data can directly be interpreted
as the {\it space-time averaged spectral function} of the $\rho$, due
to a fortuitous cancellation of the mass and $p_{T}$ dependence of the
acceptance filtering and the phase space factors associated with
thermal dilepton emission~\cite{Damjanovic:2006bd}. The two main
theoretical scenarios for the in-medium spectral properties of the
$\rho$, broadening~\cite{Rapp:1995zy} and dropping
mass~\cite{Brown:kk}, are shown for comparison. Both have been
evaluated for the same fireball evolution~\cite{Rapp:2005pr}, and the
original normalization is kept (in contrast to previous versions of
the figure~\cite{Arnaldi:2007ru,Damjanovic:2006bd}). Clearly, the
broadening scenario gets close, while the dropping mass scenario in
the version which described the CERES data reasonably
well~\cite{Rapp:1995zy,Brown:kk,Agakichiev:1995xb} fails for the much
more precise NA60 data. A strong reduction of in-medium VMD as
proposed by the vector manifestation of chiral
symmetry~\cite{Harada:2007zw} would make hadron spectral functions in
hot and dense matter altogether unobservable, but central aspects of
this scenario are totally unclear, and quantitative predictions which
could be confronted with data have not become available up to today.

\begin{figure*}[]
\begin{center}
\includegraphics*[width=0.43\textwidth]{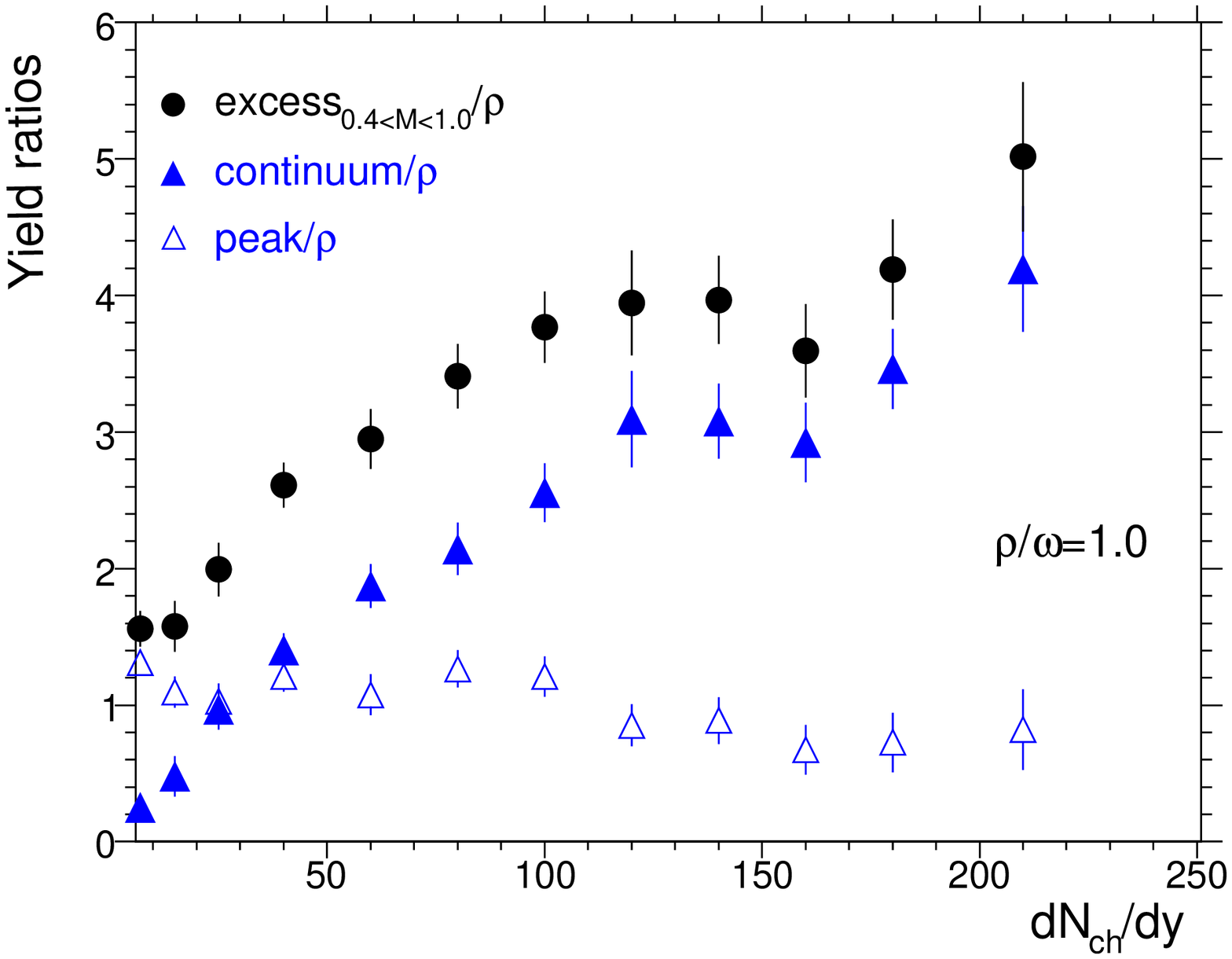}
\includegraphics*[width=0.43\textwidth]{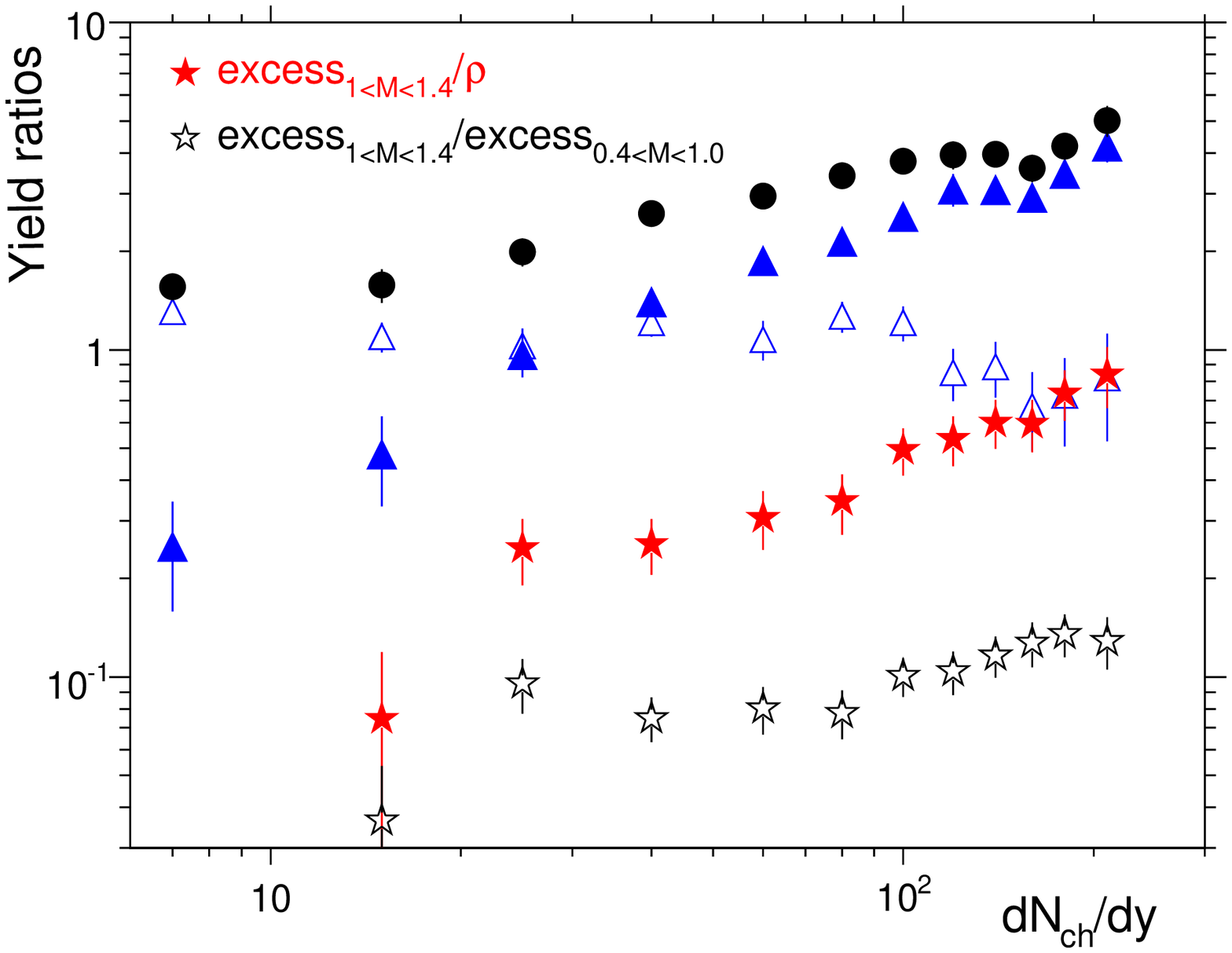}
\caption{Left: Excess yield ratios for peak, continuum and total
vs. centrality for the mass window 0.4$<$$M$$<$1 GeV. Right: same as
left on a double log scale, including here also the total in the mass
window 1.0$<$$M$$<$1.4 GeV.}
   \label{fig3}
\end{center}
\end{figure*}
A detailed view of the shape of the excess mass spectra is obtained by
using a side window method~\cite{Damjanovic:2006bd} to determine
separately the yields of the peak and the underlying continuum.  The
left panel of Fig.~\ref{fig3} shows the centrality dependence of these
variables: peak, underlying continuum and total yield in the mass
interval 0.4$<$$M$$<$1.0 GeV, all normalized to the cocktail
$\rho$. The continuum and the total show a very strong increase,
starting already in the peripheral region. In the right panel of
Fig.~\ref{fig3}, the same data of the 2$\pi$ region is plotted on a
double logarithmic scale, but here the excess in a mass window above
$M$$>$ 1GeV is also contained. This increases even steeper than the
total low-mass yield, as is clearly borne out by the rising ratio of
the two. The rise is about linear implying, in view of the
normalization to the cocktail $\rho$, that the absolute yield would be
quadratic in $N_{ch}$.

\begin{figure}[h!]
\begin{center}
\includegraphics*[width=0.43\textwidth]{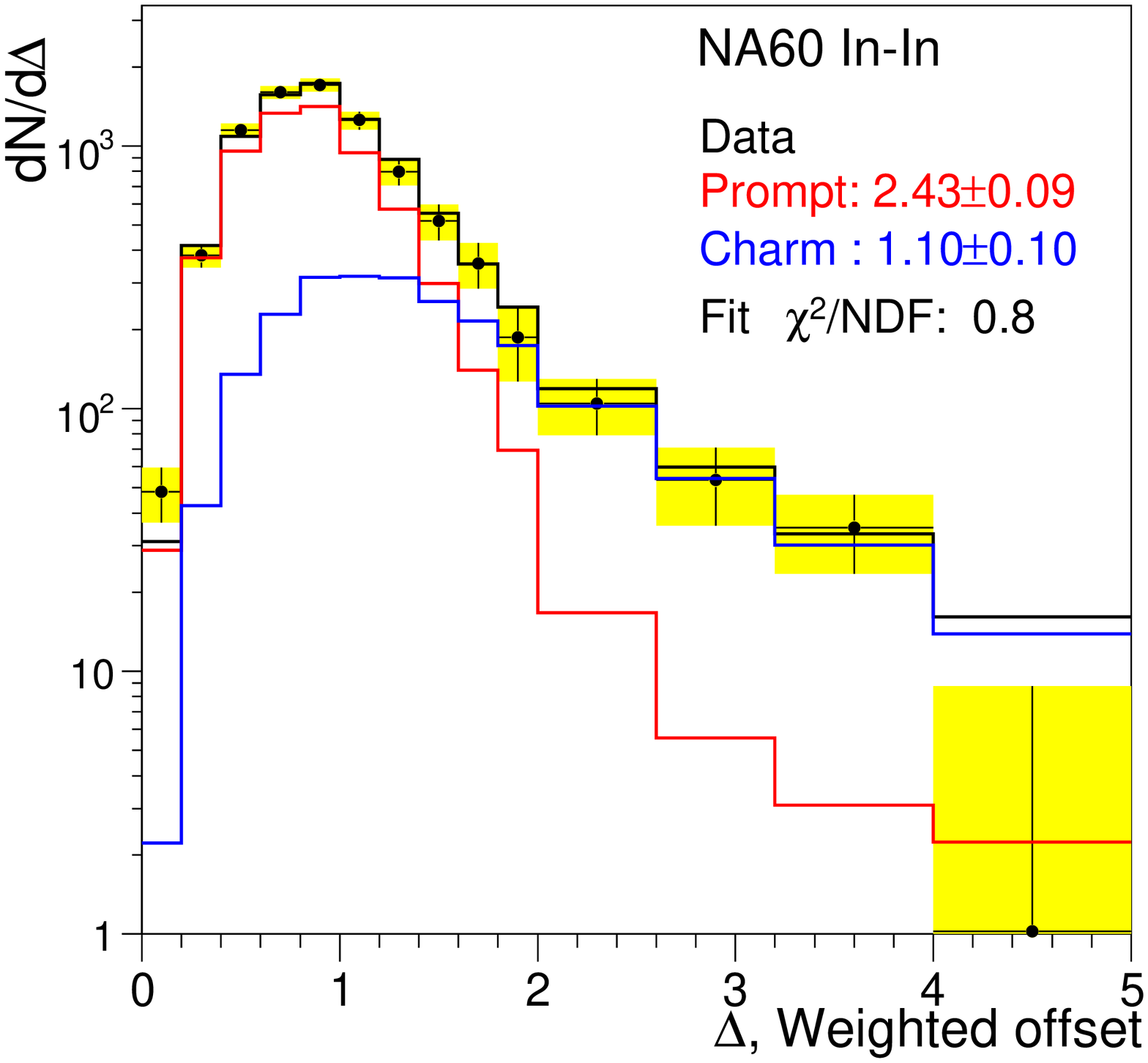}
\includegraphics*[width=0.43\textwidth]{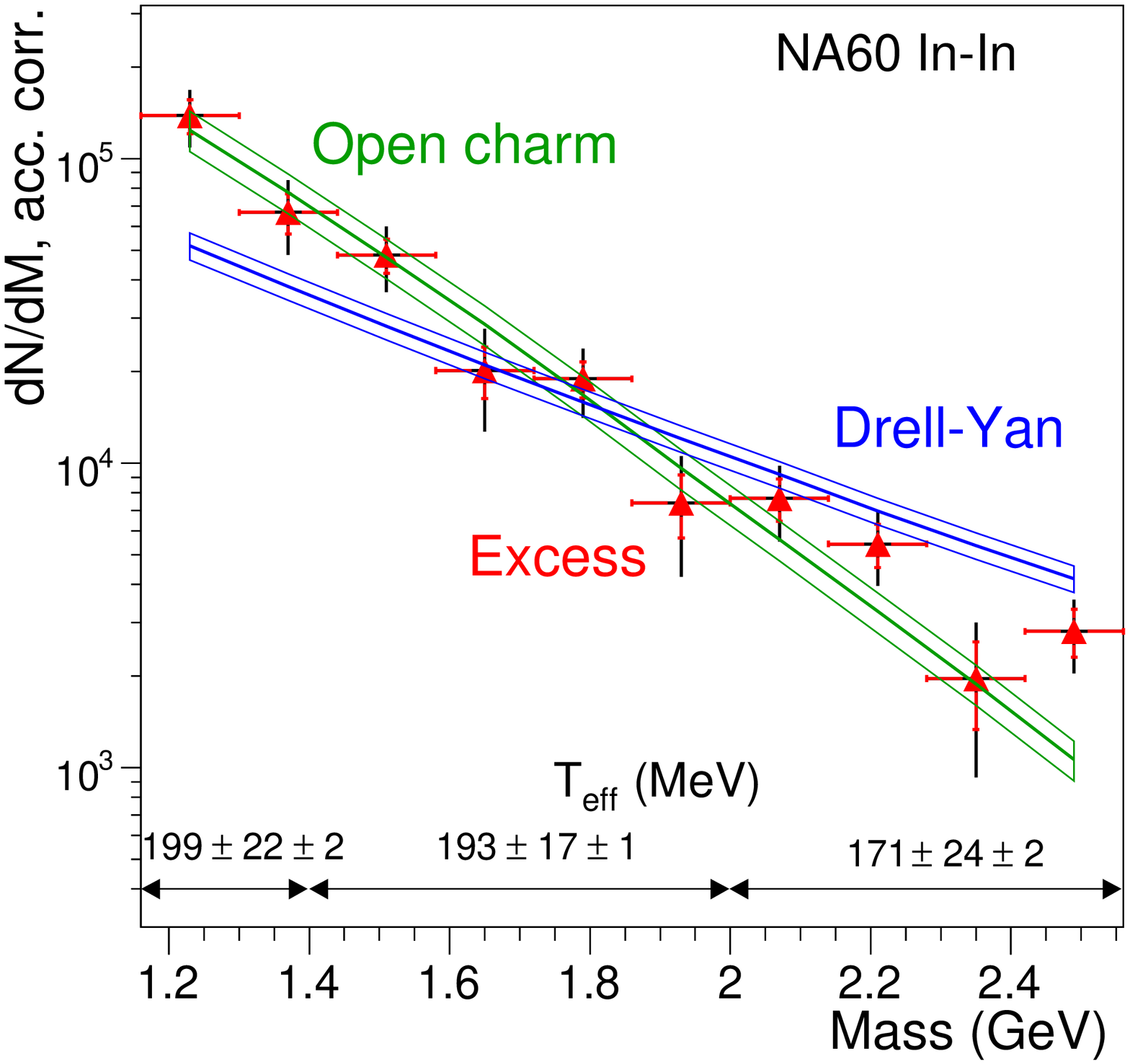}
\caption{Left: Fit of the weighted offset distribution in the IMR
region with the contributions from charm and prompt decays. Right:
Acceptance-corrected mass spectra of Drell-Yan, open charm and the
excess (triangles).}
   \label{fig4}
\end{center}
\end{figure}

The central NA60 results in the IMR region~\cite{Shahoyan:2007zz} are
shown in Fig.~\ref{fig4}. The use of the $Si$-vertex tracker allows to
measure the offset between the muon tracks and the main interaction
vertex and thereby to disentangle prompt and offset dimuons from $D$
decays. The left panel of Fig.~\ref{fig4} shows the offset
distribution to be perfectly consistent with no charm enhancement,
expressed by a fraction of $1.0\pm0.1$ of the canonical level. The
observed excess is really prompt, with an enhancement over Drell-Yan
by a factor of 2.4. The excess can now be isolated in the same way as
was done in the LMR region, subtracting the measured known sources,
here DY and open charm, from the total data. The right panel of
Fig.~\ref{fig4} shows the decomposition of the total into DY, open
charm and the prompt excess. The mass spectrum of the excess is
quite similar to the shape of open charm and much steeper than DY;
this explains of course why NA50 could describe the excess as enhanced
open charm. The transverse momentum spectra are also much steeper than
DY. The fit temperatures of the $m_{T}$ spectra associated with 3 mass
windows are indicated on the bottom of the figure.

The remainder of this paper is concerned with excess data fully
corrected for acceptance and pair
efficiencies~\cite{Arnaldi:2006jq,Damjanovic:2007qm}. In principle,
the correction requires a 4-dimensional grid in the space of
$M$-$p_{T}$-$y$-$cos{\theta_{CS}}$ (where $\theta_{CS}$ is the polar
angle of the muons in the Collins Soper frame). To avoid large
statistical errors in low-acceptance bins, it is performed instead in
2-dimensional $M$-$p_{T}$ space, using the measured $y$ and
$cos{\theta}$ distributions as an input. The latter are, in turn,
\begin{figure*}[]
\begin{center}
\includegraphics*[width=0.325\textwidth]{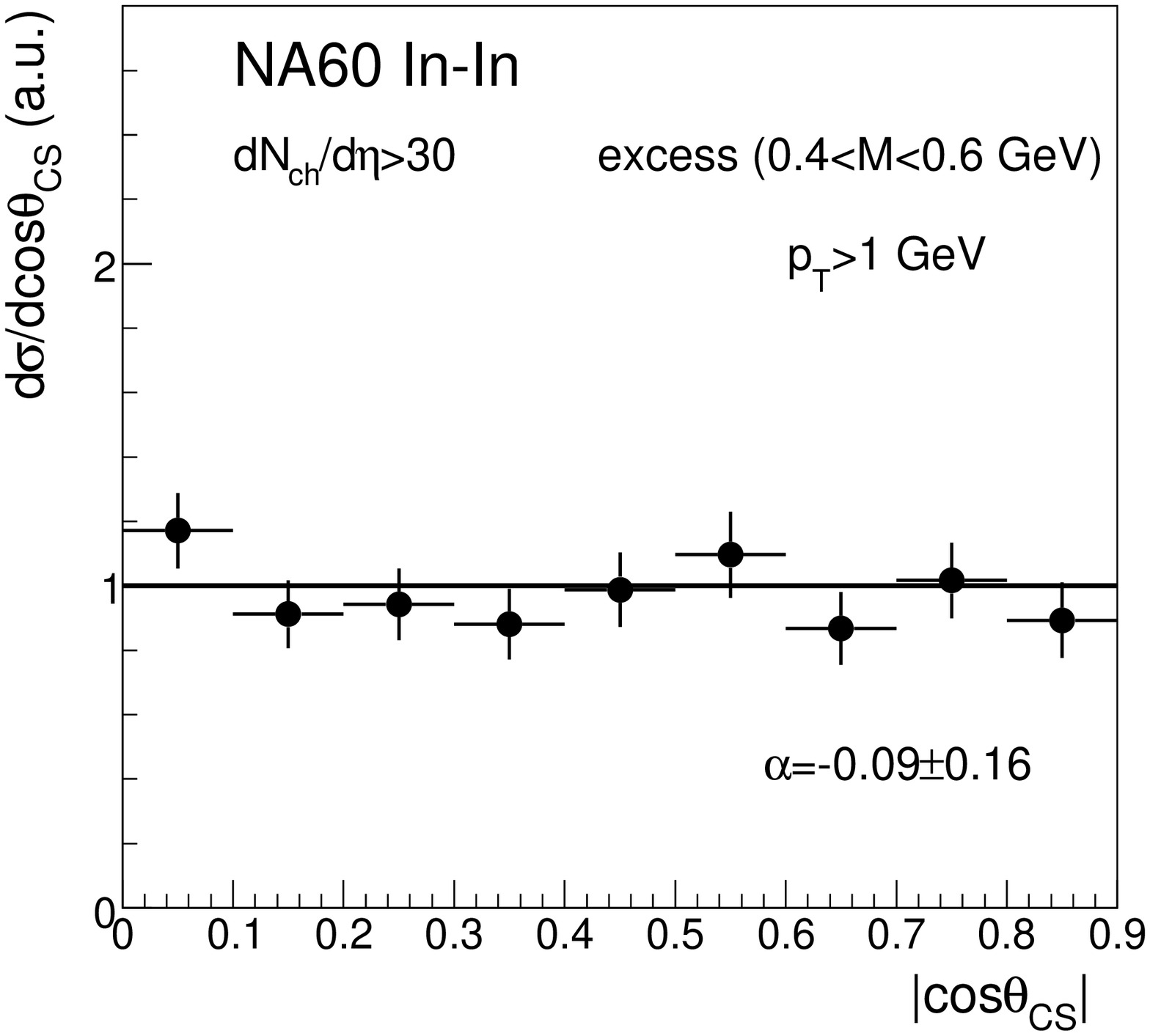}
\includegraphics*[width=0.325\textwidth]{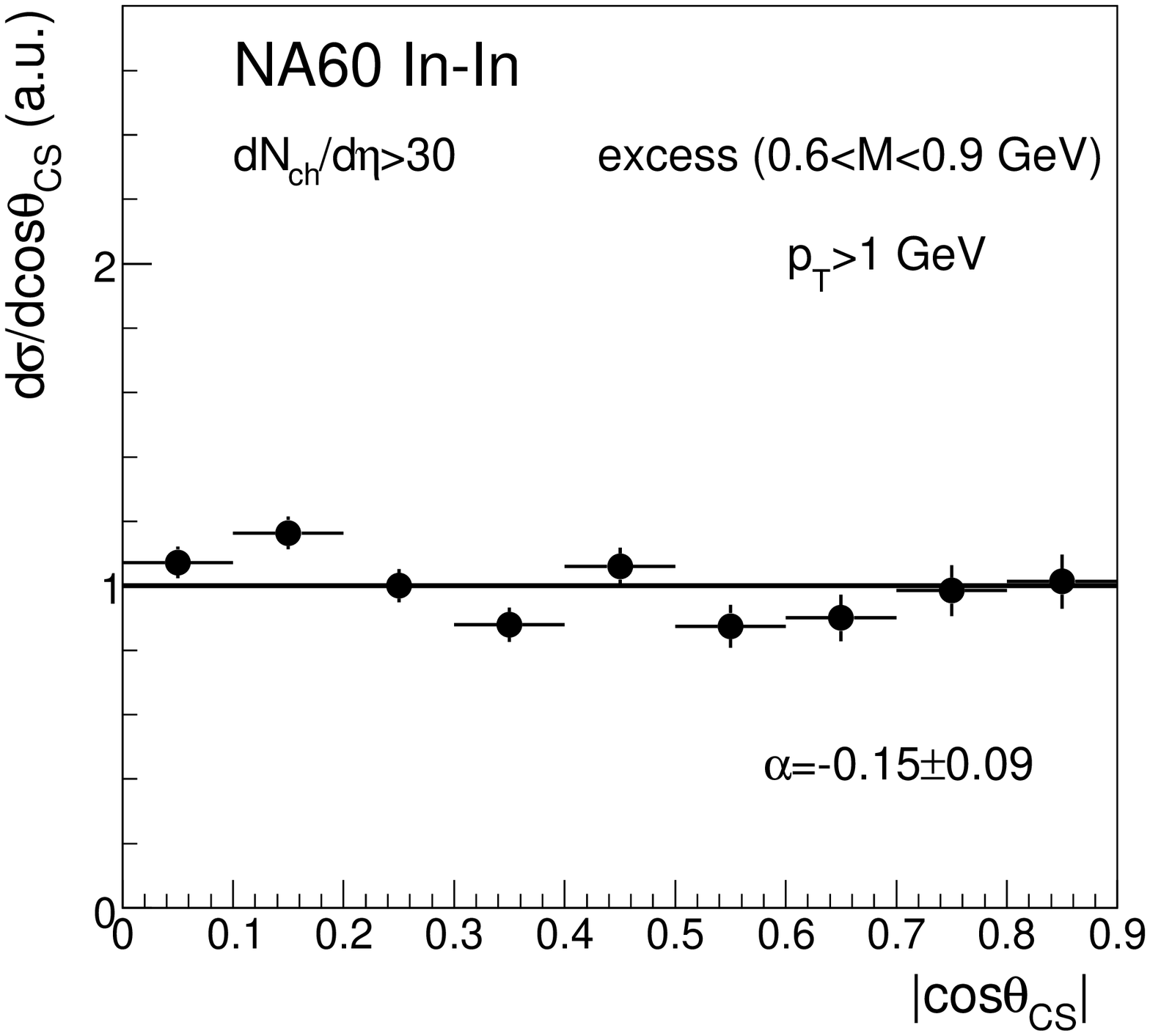}
\includegraphics*[width=0.325\textwidth]{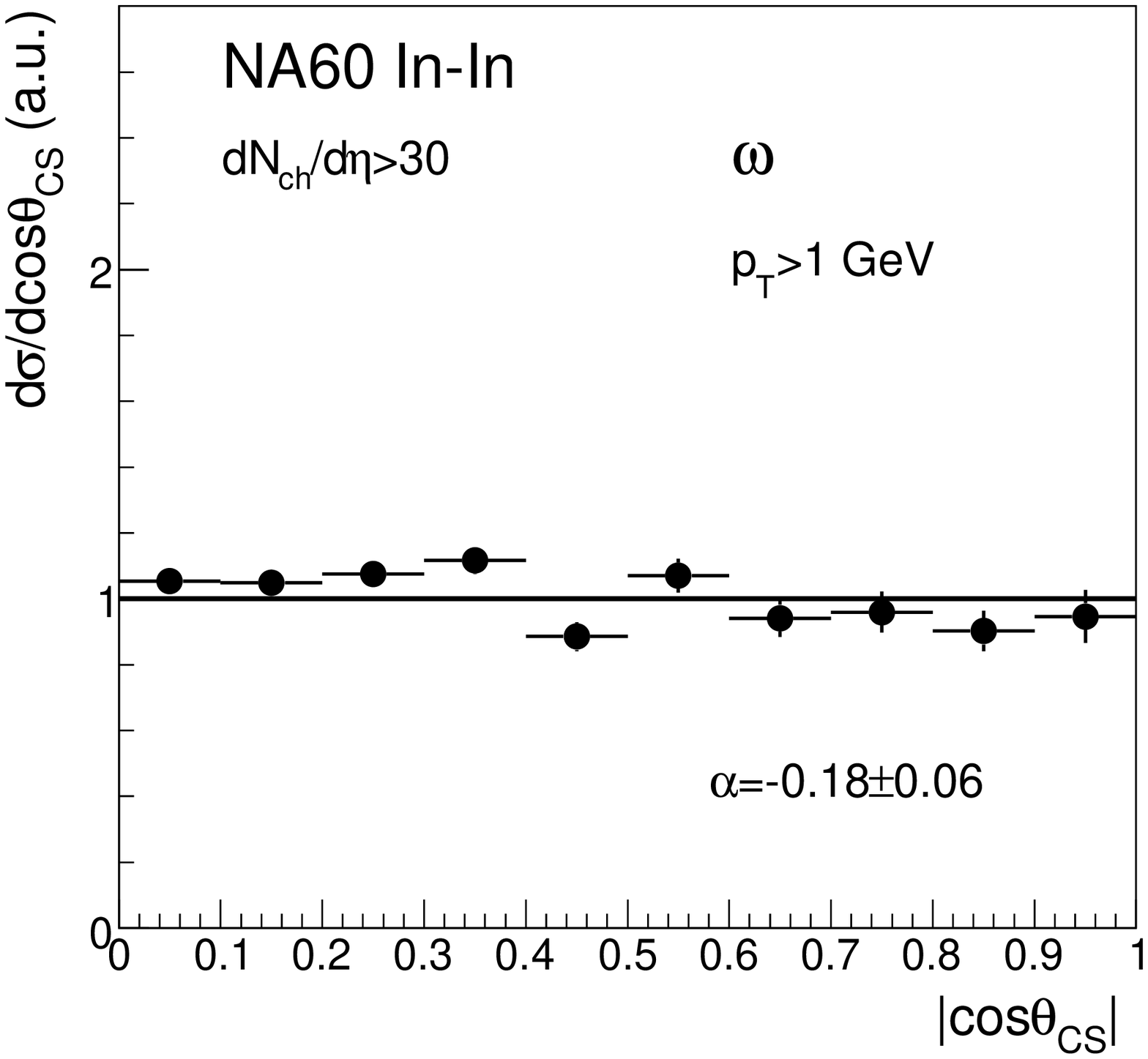}
\caption{Polar angular distributions for the excess and the $\omega$
in the Collins Soper frame.  The data are fit with the expression
$d\sigma/dcos\theta_{CS} = 1 + \alpha cos^{2}\theta_{CS}$.}
   \label{fig5}
\end{center}
\end{figure*}
obtained with acceptance corrections determined in an iterative way
from MC simulations matched to the data in $M$ and $p_{T}$. The
$y$-distribution is found to have the same rapidity width as
$dN_{ch}/d\eta$, $\sigma_{y}\sim$ 1.5~\cite{Damjanovic:2007qm}. The
$cos{\theta_{CS}}$ distributions for two mass windows of the excess
and the $\omega$ are shown in Fig.~\ref{fig5}. Within errors, they are
found to be uniform, implying the polarization of the excess dimuons
to be zero, in contrast to DY and consistent with the expectations for
thermal radiation from a randomized system.

\begin{figure*}[]
\begin{center}
\includegraphics*[width=0.43\textwidth]{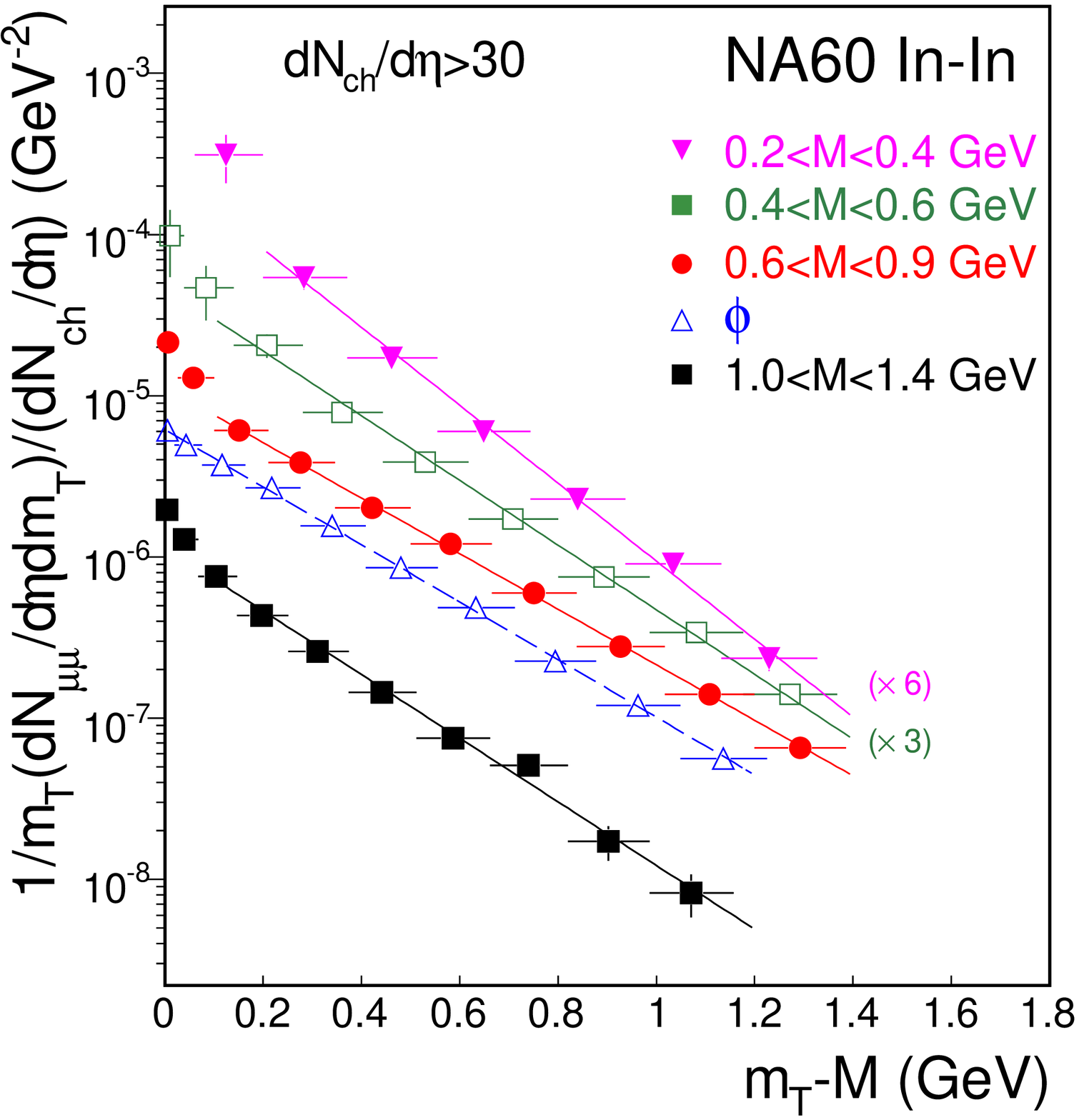}
\includegraphics*[width=0.43\textwidth]{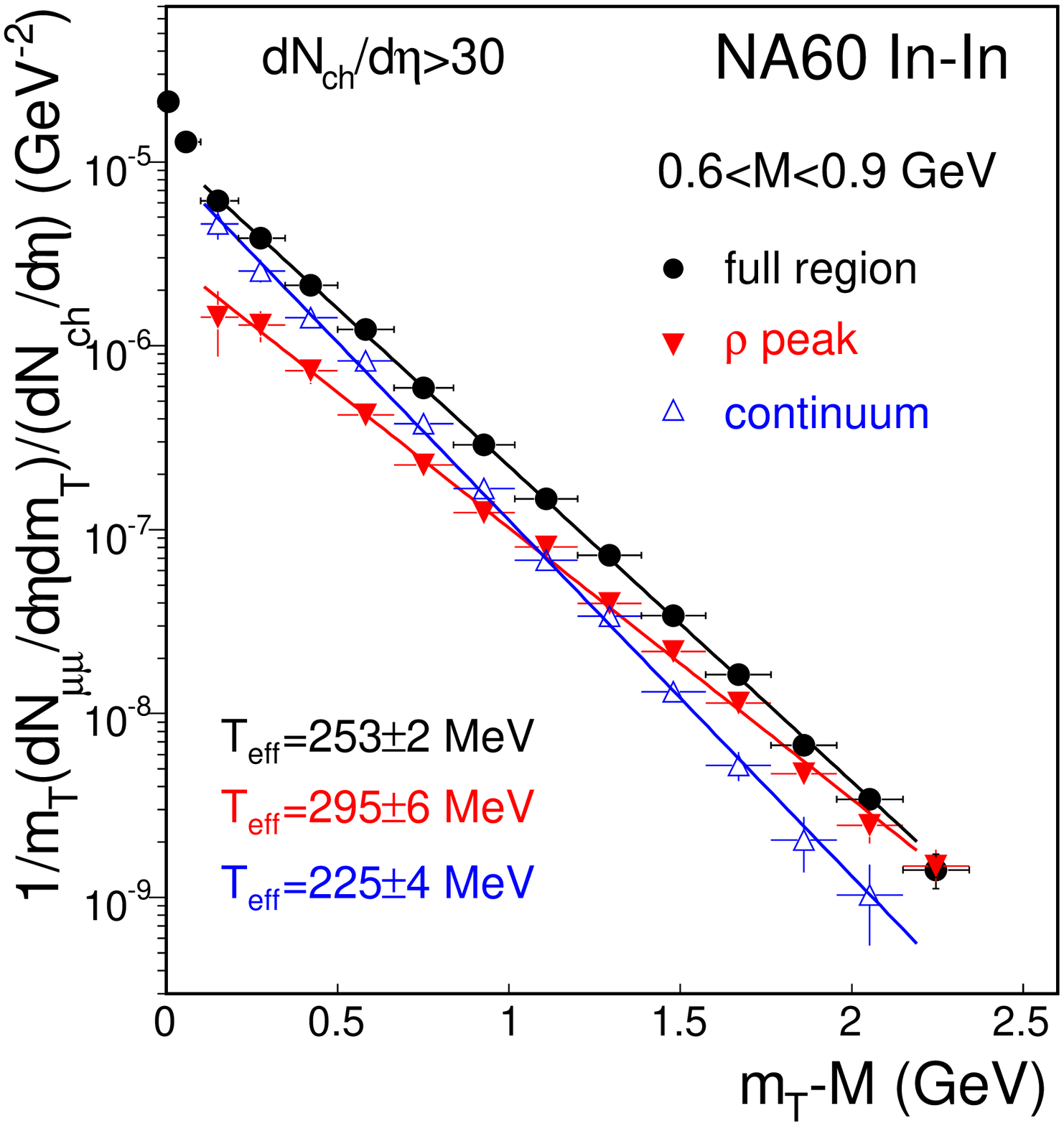}
\caption{Acceptance-corrected transverse mass spectra of the excess
dimuons for 4 mass windows and the $\phi$~\cite{Arnaldi:2006jq}
(left), and a decomposition into peak and continuum for the
$\rho$-like window (right, see text). The normalization in absolute
terms is independent of rapidity over the region measured.For error
discussion see~\cite{Arnaldi:2006jq}.}
   \label{fig6}
\end{center}
\end{figure*}

The two major variables characterizing dileptons are $M$ and $p_{T}$,
and the existence of two rather than one variable as in case of real
photons leads to much richer information. Beyond (minor) contributions
from the spectral function, $p_{T}$ encodes the key properties of the
expanding fireball, temperature and transverse (radial) flow. In
contrast to hadrons, however, which receive the full asymptotic flow
at the moment of decoupling, dileptons are continuously emitted during
the evolution, sensing the space-time development of temperature and
flow. This makes the dilepton $p_{T}$ spectra sensitive to the
emission region, providing a powerful diagnostic
tool~\cite{Kajantie:1986,Ruppert:2007cr}. Fig.~\ref{fig6} (left)
displays the centrality-integrated invariant $m_{T}$ spectra, where
$m_{T} = (p_{T}^{2} + M^{2})^{1/2}$, for four mass windows; the $\phi$
is included for comparison. The ordinate is normalized to
$dN_{ch}/d\eta$ in absolute terms, using the same procedure as
described in detail for the $\phi$~\cite{Michele:2008qm} and relating
$N_{part} \simeq dN_{ch}/d\eta$ at $\eta$=2.9 as measured to within
10\% by the $Si$ pixel telescope. Apart from a peculiar rise at low
$m_{T}$ ($<$0.2 GeV) for the excess spectra (not the $\phi$) which
only disappears for very peripheral
collisions~\cite{Specht:2007ez,Arnaldi:2006jq}, all spectra are pure
exponentials, but with a mass-dependent slope. Fig.~\ref{fig6} (right)
shows a more detailed view into the $\rho$-like mass window, using the
same side-window method as described in connection with
Fig.~\ref{fig3} to determine the $p_{T}$ spectra separately for the
$\rho$ peak and the underlying continuum. All spectra are purely
exponential up to the cut-off at $p_{T}$=3 GeV, without any signs of
an upward bend characteristic for the onset of hard processes. Their
slopes are, however, quite different (see below).

\begin{figure*}[]
\begin{center}
\includegraphics*[width=0.43\textwidth]{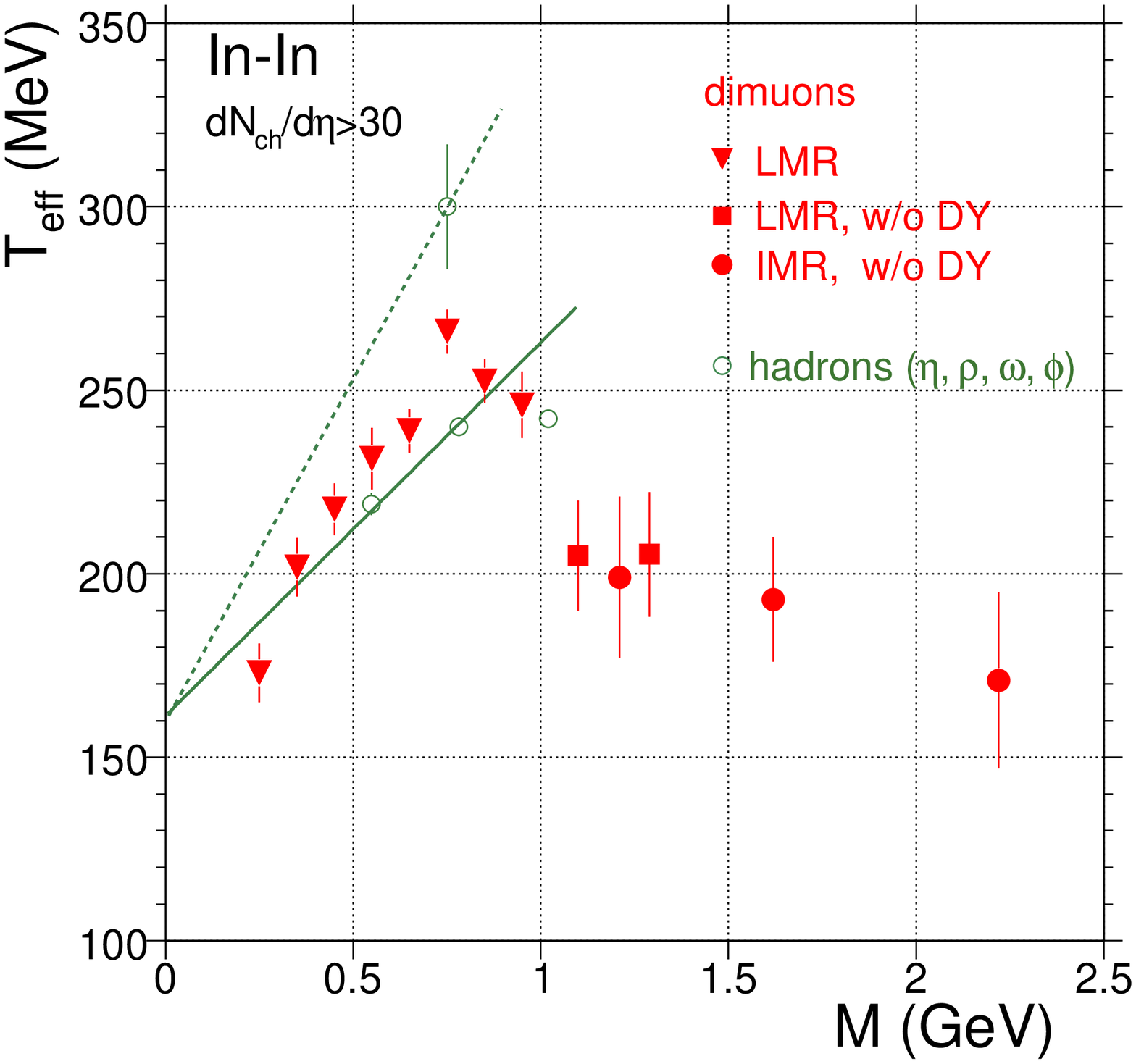}
\includegraphics*[width=0.43\textwidth]{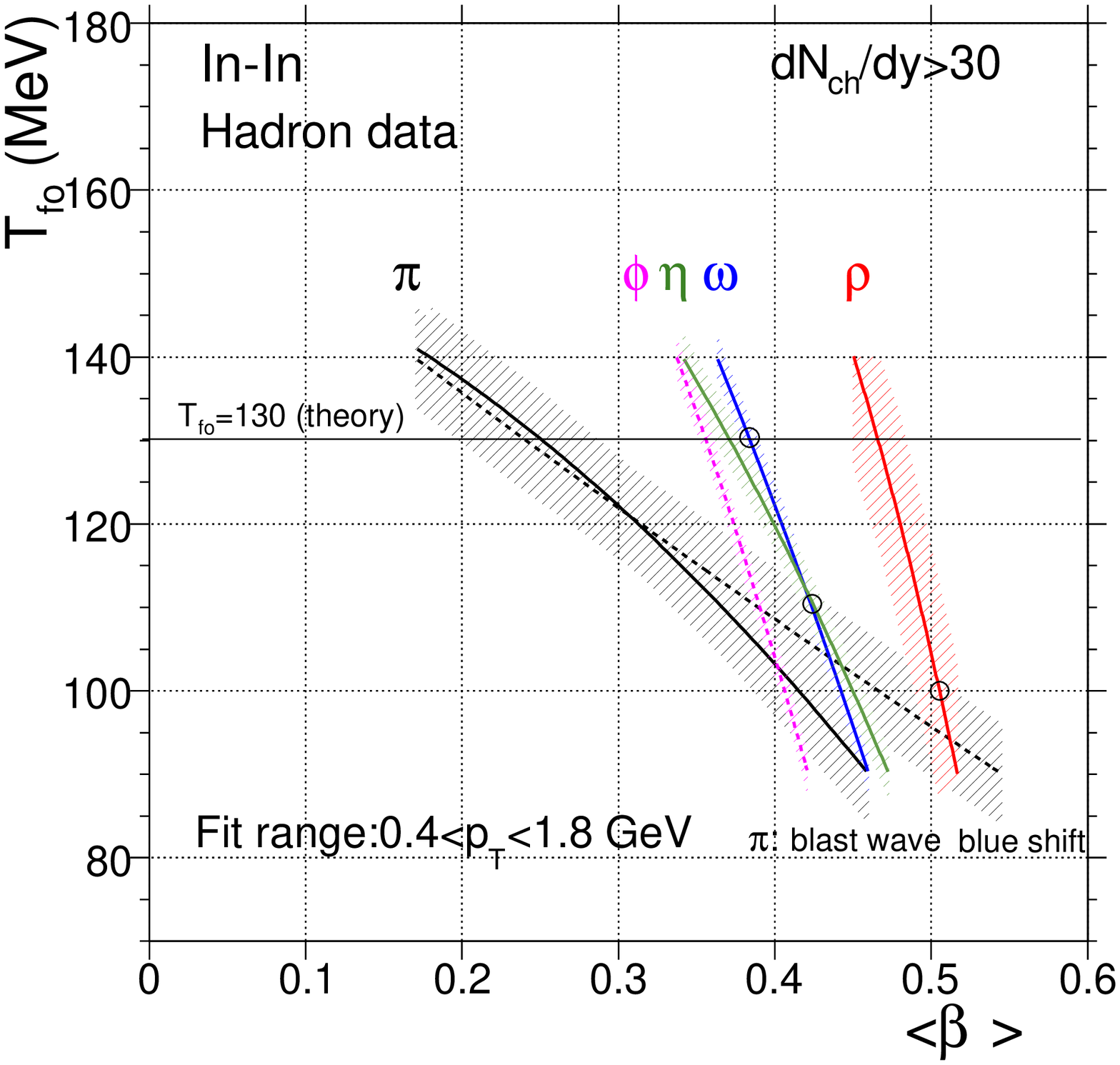}
\caption{Left: Inverse slope parameter $T_\mathrm{eff}$ vs. dimuon
mass for the combined LMR/IMR regions of the excess in comparison to
hadrons~\cite{Arnaldi:2006jq}. Open charm is subtracted
throughout. For error discussion see~\cite{Arnaldi:2006jq}. Right:
Blast wave results based on the $T_\mathrm{eff}$ values for $\pi^{-}$,
$\eta$, $\rho$, $\omega$ and $\phi$ (see text).}
   \label{fig7}
\end{center}
\end{figure*}
The inverse slope parameters $T_\mathrm{eff}$ extracted from
exponential fits to the $m_{T}$ spectra are plotted in Fig.~\ref{fig7}
(left) vs. dimuon mass~\cite{Arnaldi:2006jq}, unifying the data from
the LMR and IMR regions. The hadron data for $\eta$, $\omega$ and
$\phi$ obtained as a by-product of the cocktail subtraction are also
included, as is the single value for the $\rho$-peak from
Fig.~\ref{fig6} (right). Interpreting the latter as the freeze-out
$\rho$ without in-medium effects, all four hadron values together with
preliminary $\pi^{-}$ data from NA60 can be subjected to a simple
blast-wave analysis. The results, plotted in a plane of freeze-out
temperature $T_{f_{0}}$ and average expansion velocity $\langle
\beta_{T} \rangle$, are shown in Fig.~\ref{fig7} (right). Dilepton and
hadron data together suggest the following consistent
interpretation. Maximal flow is reached by the $\rho$, due to its
maximal coupling to pions, while all other hadrons freeze out
earlier. The $T_\mathrm{eff}$ values of the dilepton excess rise
nearly linearly with mass up to the $\rho$-pole position, but stay
always well below the $\rho$ line, exactly what would be expected for
{\it radial flow} of an {\it in-medium} {\it hadron-like} source (here
$\pi^{+}\pi^{-}$$\rightarrow$$\rho$) decaying continuously into dileptons.

Beyond the $\rho$, however, the $T_\mathrm{eff}$ values of the excess
show a sudden decline by about 50 MeV. Extrapolating the lower-mass
trend to beyond the $\rho$, such a fast transition to a seeming
low-flow situation is extremely hard to reconcile with emission
sources which continue to be of dominantly hadronic origin in this
region. A more natural explanation would then be a transition to a
dominantly early {\it partonic} source with processes like
$q\bar{q}\rightarrow \mu^{+}\mu^{-}$ for which flow has not yet built
up~\cite{Ruppert:2007cr}. While still
controversial~\cite{vanHees:2006ng}, this may well represent the first
direct evidence for thermal radiation of partonic origin, overcoming
parton-hadron duality for the {\it yield} description in the mass
domain.

\begin{figure*}[]
\begin{center}
\includegraphics*[width=0.29\textwidth]{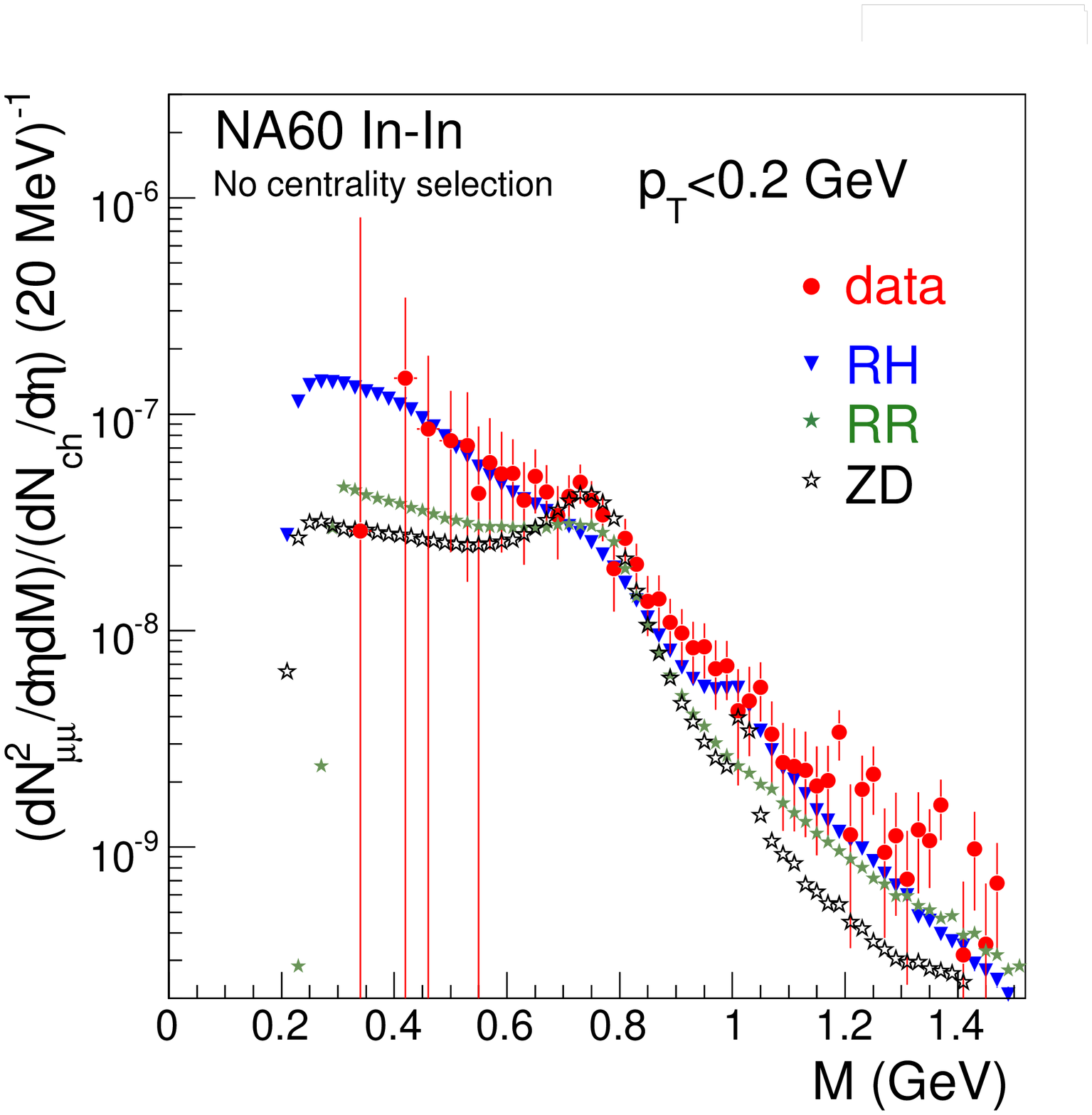}
\includegraphics*[width=0.29\textwidth]{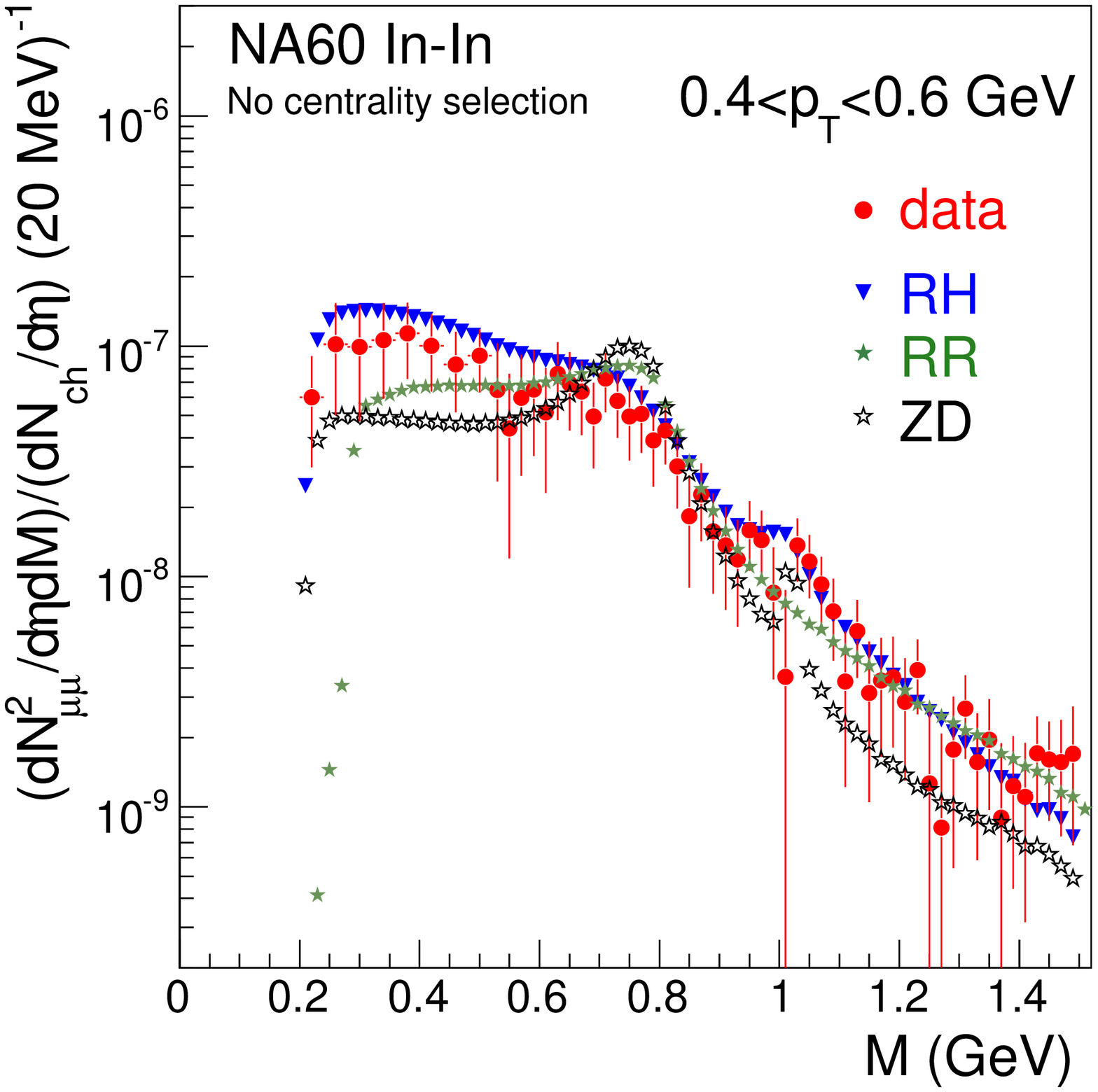}
\includegraphics*[width=0.29\textwidth]{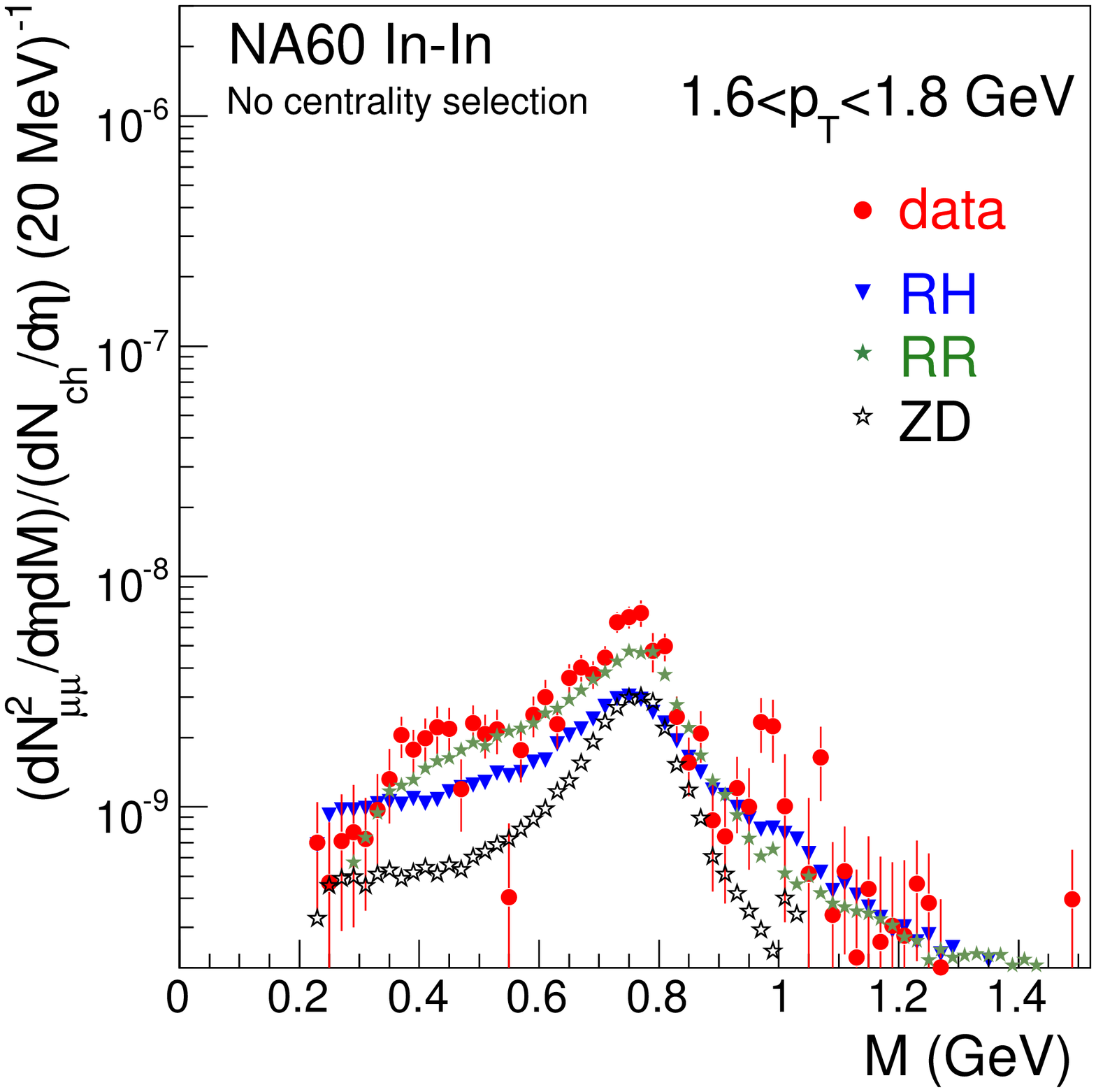}
\includegraphics*[width=0.29\textwidth]{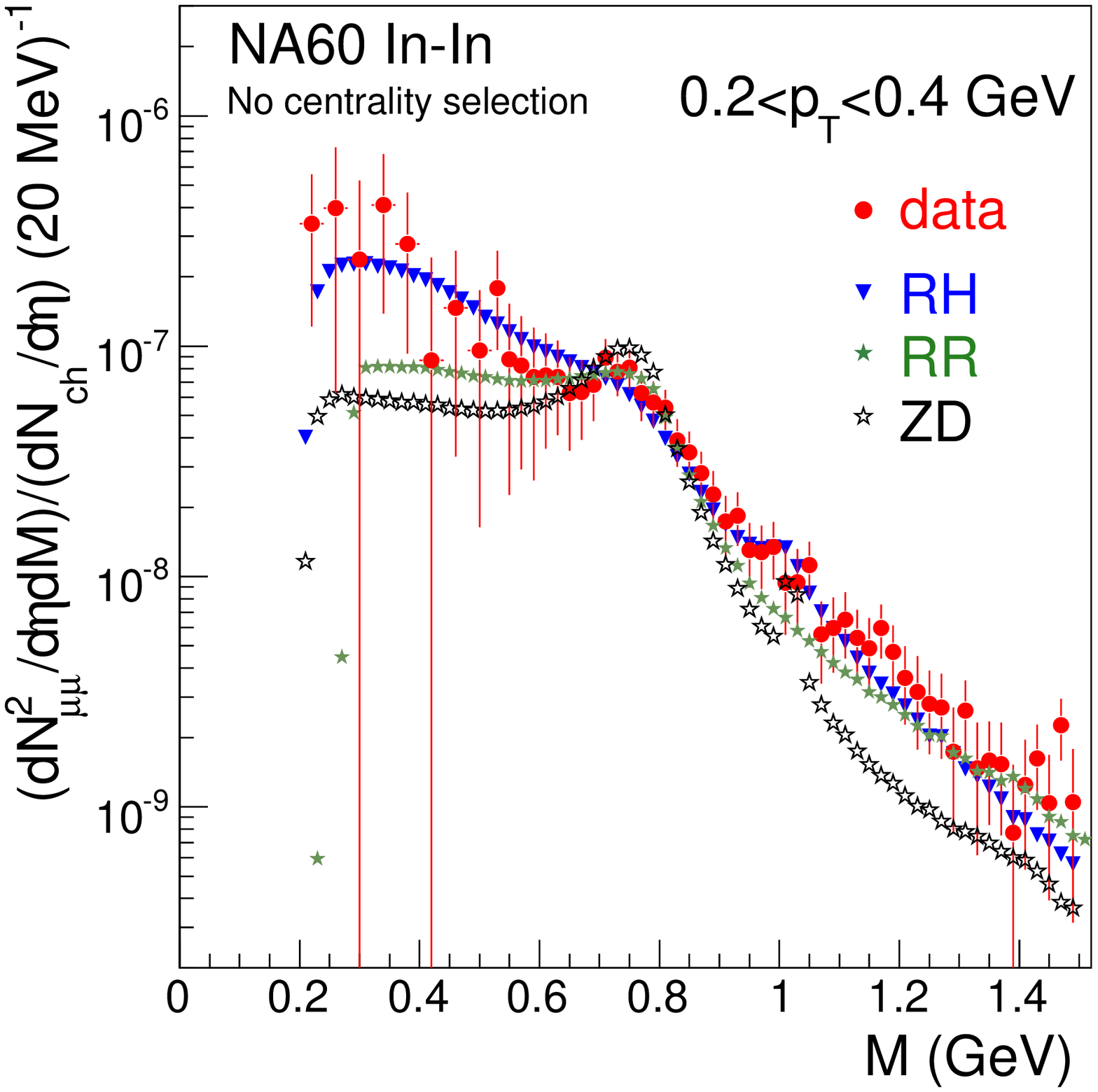}
\includegraphics*[width=0.29\textwidth]{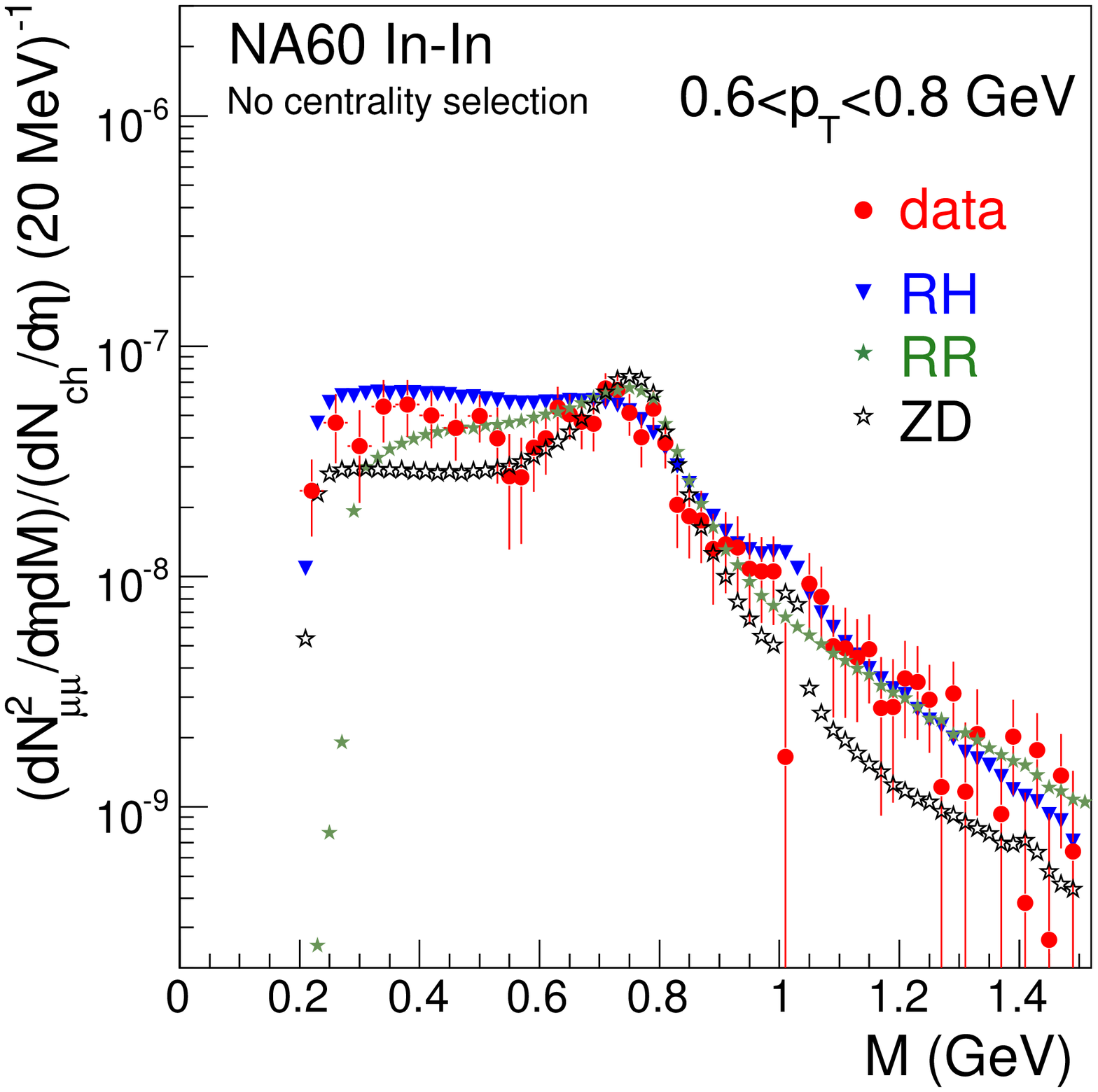}
\includegraphics*[width=0.29\textwidth]{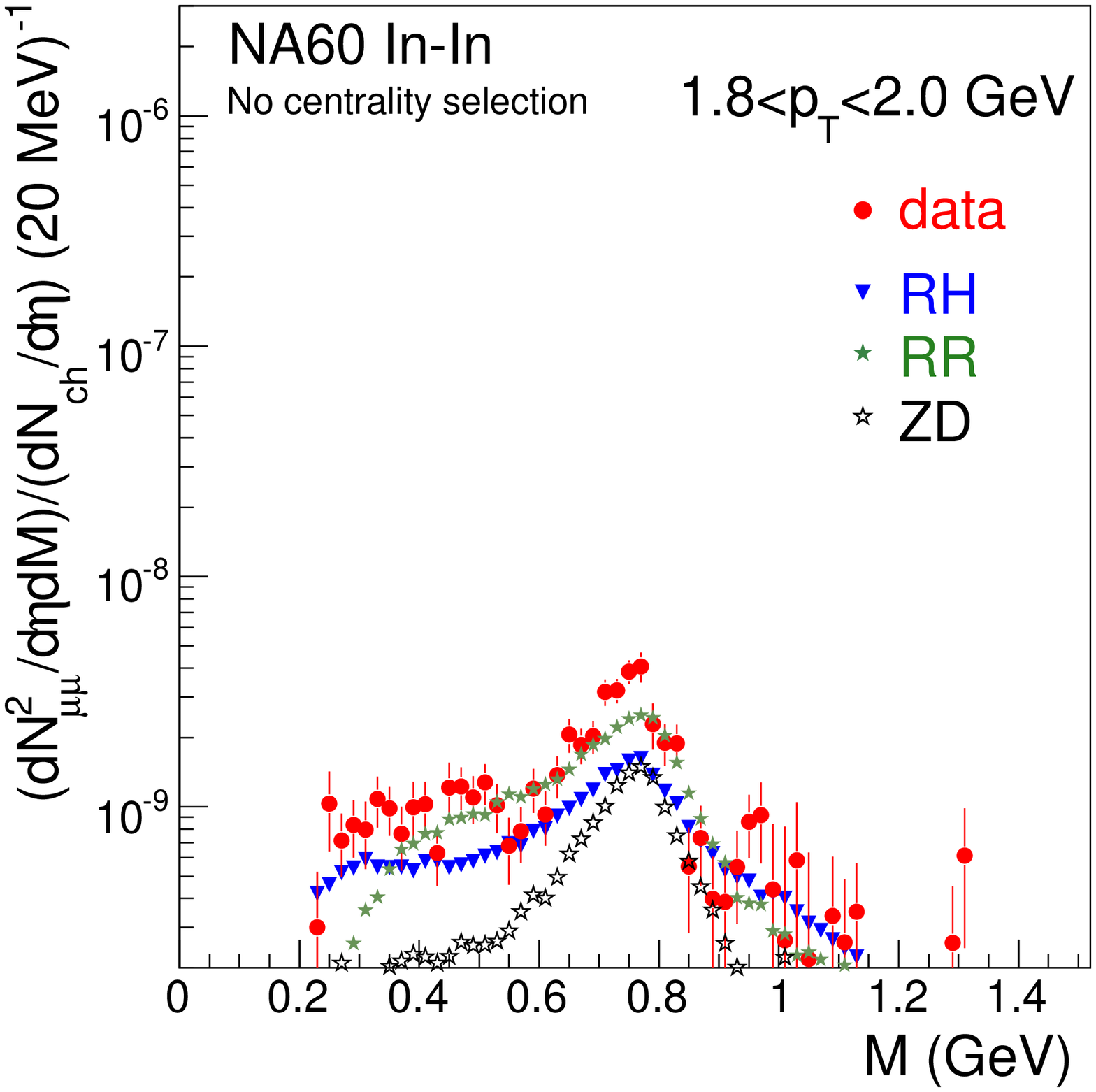}
\caption{Acceptance-corrected mass spectra of the excess dimuons in
selected slices of $p_{T}$. Absolute normalization as in
Fig.~\ref{fig6}.}
   \label{fig8}
\end{center}
\end{figure*}

The acceptance- and efficiency-corrected data can also be projected on
the mass axis. Fig.~\ref{fig8} shows a set of mass spectra for some
selected slices in $p_{T}$ to illustrate the evolution from low to
high $p_{T}$. Recent theoretical results from the three major groups
working in the field are included for
comparison~\cite{Ruppert:2007cr,vanHees:2006ng,Dusling:2007rh}. At
very low $p_{T}$, a strong rise towards low masses is seen in the
data, reflecting the Boltzmann factor, i.e. the Plank-like radiation
associated with a very broad, nearly flat spectral function. Only the
Hees/Rapp scenario~\cite{vanHees:2006ng} is able to describe this part
quantitatively, due to their particularly large contribution from
baryonic interactions to the low-mass tail of the $\rho$ spectral
function. At higher $p_{T}$, the influence of radial flow increasingly
changes the spectral shapes, and at very high $p_{T}$, all spectra
appear $\rho$-like. Here, only the Renk/Ruppert
results~\cite{Ruppert:2007cr} seem to contain sufficient flow to
describe the data.

This paper contains the most comprehensive data set on excess
dileptons above the known sources which has so far become available
through NA60. All observations can consistently be interpreted in
terms of thermal radiation from the fireball in the whole mass region
up to the $J/\psi$. The superior data quality and the resulting
clarity of the physics messages has yet to be matched by any other
dilepton experiment in the field.

\end{document}